\documentclass[twocolumn,prb,showpacs]{revtex4}
\usepackage{graphicx}
\usepackage{amssymb}
\usepackage{bm}
\newcommand{\GTO}{$\rm Gd_2Ti_2O_7$}
\newcommand{\GGG}{$\rm Gd_3Ga_5O_{12}$}
\begin{document}

\title{Magnetocaloric effect in a pyrochlore
antiferromagnet \GTO}

\author{S. S. Sosin, L. A. Prozorova, A. I. Smirnov}
\affiliation{P. L. Kapitza Institute for Physical Problems RAS,
119334 Moscow, Russia}
\author{A. I. Golov, I. B. Berkutov}
\altaffiliation[present address:\ ]{Institute for Low Temperature
Physics and Engineering, 47, Lenin av., 61103 Kharkov, Ukraine}
\affiliation{Department of Physics and Astronomy, University of
Manchester, Manchester M13 9PL, UK}
\author{O. A. Petrenko, G. Balakrishnan}
\affiliation{Department of Physics, University of Warwick,
Coventry CV4 7AL, UK}
\author{M. E. Zhitomirsky}
\affiliation{Commissariat \'{a} l'Energie Atomique,
DSM/DRFMC/SPSMS, 38054 Grenoble, France}

\date{October 21, 2004}

\begin{abstract}
An adiabatic demagnetization process is studied in \GTO, a
geometrically frustrated antiferromagnet on a pyrochlore lattice.
In contrast to conventional paramagnetic salts, this compound can
exhibit a temperature decrease by a factor of ten in the
temperature range below the Curie-Weiss constant. The most
efficient cooling is observed in the field interval between 120
and 60~kOe corresponding to a crossover between saturated and
spin-liquid phases. This phenomenon indicates that a considerable
part of the magnetic entropy survives in the strongly correlated
state. According to the theoretical model, this entropy is
associated with a macroscopic number of local modes remaining
gapless till the saturation field. Monte Carlo simulations on a
classical spin model demonstrate good agreement with the
experiment. The cooling power of the process is experimentally
estimated with a view to possible technical applications. The
results for \GTO\ are compared to those for \GGG, a well-known
material for low temperature magnetic refrigeration.
\end{abstract}


\maketitle

\section{Introduction}

The distinct feature of highly frustrated magnetic materials is a
peculiar spatial arrangement of the magnetic ions.
Antiferromagnets on typical geometrically frustrated structures,
like, for instance, kagome, garnet, and pyrochlore lattices, have
an infinite number of classical ground states. This macroscopic
degeneracy precludes any type of conventional magnetic ordering.
As a result, frustrated magnets remain in a disordered cooperative
paramagnetic ground state at temperatures well below the
paramagnetic Curie-Weiss constant
$\theta_{CW}$.\cite{villain,chalker,moessner,canals} Weak residual
interactions or quantum and thermal fluctuations usually induce
some kind of ordered or spin-glass state at $T^*\ll\theta_{CW}$. A
number of geometrically frustrated magnets have been
experimentally studied in the past decade.\cite{ramirezR} Magnetic
pyrochlore compound \GTO\ is one of the prototype examples. The
Gd$^{3+}$ ions have spin $S=7/2$ and zero orbital momentum, which
yields a good realization of a nearest-neighbor Heisenberg
exchange antiferromagnet on a pyrochlore lattice. Recent specific
heat, susceptibility and neutron scattering measurements
\cite{raju,champion,ramirez02} have shown that \GTO\ remains
disordered over a wide temperature interval below its
$\theta_{CW}\simeq 10$~K. The transition to an ordered phase is
presumably driven by weak dipole-dipole interactions and occurs
only at $T_{N1}\approx 1$~K.

Infinite degeneracy of the magnetic ground state of a frustrated
magnet implies the presence of a macroscopic number of local
zero-energy modes. Such soft modes correspond to rotational
degrees of freedom of a finite number of spins with no change in
the total exchange energy. In a pyrochlore structure (a matrix of
corner-sharing tetrahedra on the fcc lattice), these groups of
spins are hexagons formed by the edges of neighboring tetrahedra,
which lie in the kagome planes [(111) and equivalent planes] (see
{\it e.g.} Ref.~\onlinecite{broholm}). In zero applied field, if
the six spins are arranged antiferromagnetically around one
hexagon, they effectively decouple from the other spins and can
rotate by an arbitrary angle. The low-energy hexagon modes have
been observed in quasielastic neutron scattering studies on spinel
compound ZnCr$_2$O$_4$, which is a spin-3/2 Heisenberg
antiferromagnet on a pyrochlore lattice.\cite{broholm} The
thermodynamic consequence of these local excitations is that a
considerable fraction of the magnetic entropy is not frozen down
to temperatures much less than $\theta_{CW}$.

An interesting effect related to a field evolution of the
zero-energy local modes is an enhanced magnetocaloric effect near
the saturation field $H_{\rm sat}$ predicted in
Ref.~\onlinecite{mzh03}. Transformation between a nondegenerate
fully polarized spin state above $H_{\rm sat}$ and an infinitely
degenerate magnetic state below $H_{\rm sat}$ occurs via
condensation of a macroscopic number of local modes and produces
large changes of entropy in magnetic field. Quantum fluctuations do
not destroy such an effect.\cite{mzh04} In particular, at
$H=H_{\rm sat}$ geometrically frustrated magnets on pyrochlore and
kagome lattices have a finite macroscopic entropy, which does not
depend on the value of the spin of the magnetic ions. The finite entropy of
kagome antiferromagnet has been calculated exactly,\cite{mzh04}
whereas the corresponding quantity for a pyrochlore lattice remains
unknown. Therefore, an experimental observation of the
magnetocaloric effect in \GTO\ should be important both from a
fundamental point of view, as a physical probe of the local
zero-energy excitations in the frustrated ground state below
$H_{\rm sat}$, and because of the possible technological
applications. The enhanced magnetic cooling power of gadolinium
gallium garnet \GGG\ has been known for a long time,
\cite{fisher,barclay,daudin} although without reference to the
frustrated nature of a garnet (hyper-kagome) lattice. Experiments
have been also performed on other garnets.\cite{numazava} Thus, a
comparison between the two typical frustrated magnets is important
since a stronger frustration on a pyrochlore lattice produces the
maximal cooling rate among all known types of geometrically
frustrated magnets.\cite{mzh03} In the present work we study
experimentally and theoretically the magnetocaloric effect in
\GTO. Demagnetization of a \GTO\ sample under quasi-adiabatic
conditions is performed and compared to a similar study of \GGG.
Classical Monte Carlo simulations of an ideal adiabatic
demagnetization process are performed for both materials. Entropy
variations and the cooling power of \GTO\ in an applied magnetic
field are calculated from a combination of the adiabatic
demagnetization results and specific heat measurements.

\section{Experiment}
\subsection{Specific heat measurements}

\begin{figure}[h]
\centerline{\includegraphics[width=\columnwidth]{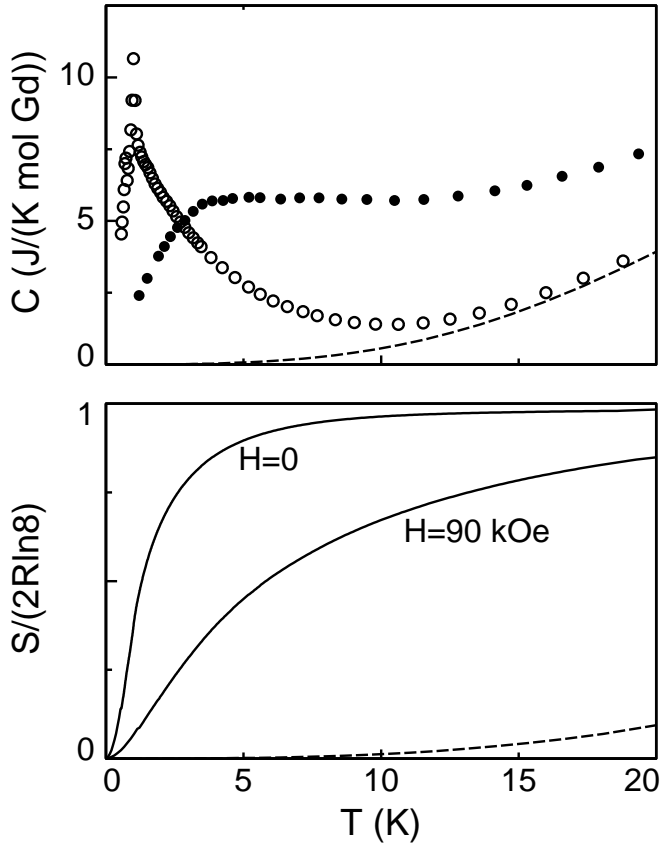}}
\caption{Temperature dependence of the specific heat (upper panel)
and the entropy (lower panel) of the \GTO\ single crystal sample:
{\Large $\circ$} -- zero field, {\Large $\bullet$} -- $H=90$ kOe;
solid lines are the results of $C/T$ integration, dashed lines
represent the phonon contribution.} \label{fig1}
\end{figure}

A single-crystal sample of \GTO\ was grown by the method described
in Ref.~\onlinecite{oleg98} and is approximately $3.5\times
1.5\times 1$ mm$^3$ in size (32.5 mg by mass). For preliminary
estimation, we present the results of specific heat measurements
of \GTO\ performed in a Quantum Design PPMS calorimeter at zero
field and at $H=90$~kOe (see Fig. \ref{fig1}, upper panel). The
curve at $H=0$ was taken from our previous work in Ref.
\onlinecite{oleg03} while the high field data were measured in the
temperature range from 1.5 to 20~K. Instead of two sharp peaks on
a zero field $C(T)$ dependence at $T_{N1}=1$ K and $T_{N2}=0.75$ K
corresponding to the ordering phase transitions, the high field
curve demonstrates a broad maximum around 5--6 K. The lattice
contribution to the specific heat extracted by fitting the zero
field data at temperatures 30--70 K in the Debye approximation is
shown in the figure by a dashed line. The temperature dependence
of the magnetic entropy (total entropy with the phonon part
subtracted) obtained by integrating $C/T$ curves are presented at
the lower panel of Fig.~1. They give a general overview of the
magnetocaloric effect in this compound. An adiabatic
demagnetization of the system down to zero field starting at
$H=90$ kOe and $T=10$ K (around $\theta_{CW}$) results in
decreasing temperature to 2 K, while the isothermal process at
$T=2$ K is accompanied by an entropy increase by approximately one
half of the total magnetic entropy.

\subsection{Adiabatic demagnetization}

In the main part of the experiment, the magnetocaloric effect in
\GTO\ is studied in detail by measuring the temperature of the
sample in a quasi-adiabatic regime as a function of time and
magnetic field. For this purpose, a commercially available
thin-film RuO$_2$ resistor with the resistance calibrated down to
100~mK in fields up to 120~kOe was glued onto the sample. It was
also used as a heater to regulate the starting temperature of the
experiment. The sample was suspended on four thin constantan wires
($20 \mu$m in diameter and 5~cm in length) soldered to the
thermometer to make a 4-wire resistance measurement. The
experimental cell was put in a vacuum can immersed in a helium
bath held at 1.8--4.2~K. The heat exchange gas inside the can was
absorbed by a charcoal cryopump to a pressure of $10^{-7}$~torr.
The cryopump was equipped with a heater to desorb some exchange
gas when necessary to cool the sample during the experiment.
Magnetic fields up to 120~kOe were generated by a superconducting
magnet. The field $H$ was applied perpendicular to the $\langle
111\rangle$ axis. No correction for the demagnetization factor was
made. The measurements were done at a sweep rate of 10~kOe/min.

\begin{figure}[t]
\centerline{\includegraphics[width=\columnwidth]{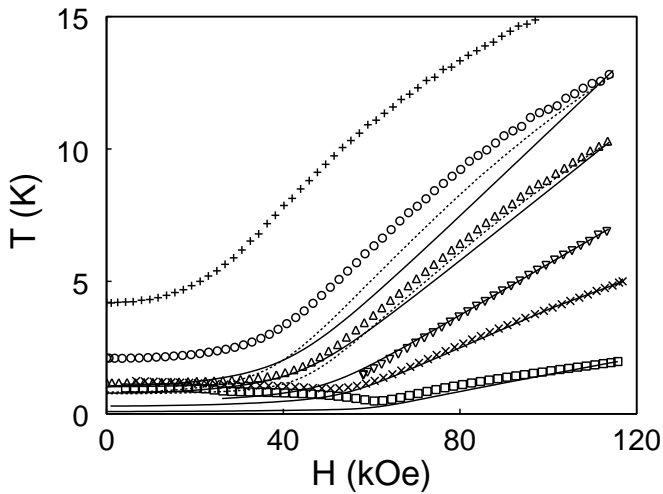}}
\caption{Temperature variations of the adiabatic demagnetization
of \GTO for various starting temperatures; dotted lines represent
the corrections for the lattice heat capacity (see Discussion).
Solid lines are obtained by Monte Carlo simulations with the
exchange constant $J=0.3$~K.} \label{fig2}
\end{figure}

The temperature versus magnetic field curves for different
starting temperatures are shown in Fig.~\ref{fig2}. The
temperature drop on demagnetization $T_i/T_f$ depends on the
starting temperature. It has a maximum around $T_i\simeq 10$~K
where the temperature drops by a factor of 10. A characteristic
feature for all the $T_S(H)$ curves starting below 10~K is that a
great deal of cooling occurs in the field range from 120 to
60~kOe, which contrasts sharply with a continuous adiabatic
cooling ($T/H={\rm const}$) of an ideal paramagnet.

\begin{figure}
\centerline{\includegraphics[width=\columnwidth]{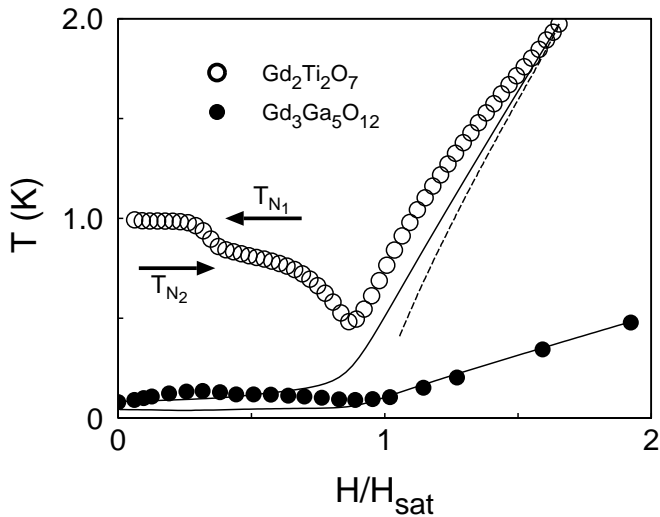}}
\caption{Comparison of adiabatic demagnetization in \GTO\ (raw
data) and \GGG\ on a field scale normalized to $H_{\rm sat}$.
Monte Carlo simulations are shown by the solid lines; the dashed
line is an overcorrection to the experimental data made using
formula (\ref{Tcor}).} \label{fig3}
\end{figure}

Additionally, a parasitic temperature drift $\dot{T}_p$ (due to
radiation, wiring and the measuring current) was investigated at
several fixed values of the field for temperatures along each
curve $T_S(H)$. The values of $\dot{T}_p$ have a smooth
temperature dependence within $\pm 2$ mK/sec changing sign at
about 10 K. The experimental data were corrected at each moment of
time $t_0$ (corresponding to each value of the magnetic field $H$)
using a set of measured $\dot{T}_p$ values by formula
\begin{equation}
T_{\rm cor}(H,t_0) = T_{\rm exp}(H,t_0)-\int_0^{t_0}\dot{T}_p dt \
, \label{Tcor}
\end{equation}
where $T_{\rm exp}$ is the raw experimental value, $T_{\rm cor}$
is the corrected temperature. One should mention, that the above
correction procedure is applicable only if the parasitic heating
does not strongly affect the whole demagnetization process. In
this case, after the correction is applied to the raw data, the
process appeared to be fully reversible by the sweep direction, so
that the curves obtained on increasing the magnetic field do not
differ from those shown in Fig.~\ref{fig2}. A more general
heat-balance equation
\begin{equation}
W_p dt =  C dT + T\left(\partial S/\partial H\right)_T dH \ ,
\label{balance}
\end{equation}
($W_p$ is a parasitic heat leak ($\dot T_p=W_p/C$), $C$ is the
specific heat at a constant field) leads to formula (\ref{Tcor})
if the function $T(\partial S/\partial H)_T/C \equiv -(\partial
T/\partial H)_S$ does not change significantly between $T_{\rm
exp}$ and $T_{\rm cor}$. Such an assumption is valid for all the
scans shown in Fig. \ref{fig2} except the one starting at $T_i=2$
K and the corresponding $T_S(H)$ curves are given with the
appropriate subtraction. For the lowest scan, the correction
appears to be of the same order of magnitude with the experimental
temperature. In addition, the derivative $(\partial T/\partial
H)_S$ strongly decreases in the vicinity of the ordering
transition which leads to an overcorrection of the experimental
data. This overcorrection is illustrated in Fig.~\ref{fig3} by the
dashed line. Thus, an ideal adiabatic dependence $T_S(H)$ should
lie between the raw experimental data and the corrected curve. The
minimum temperature reached experimentally on demagnetization from
$T_i=2$~K at $H_f=62$~kOe is $T_{\rm min}=0.48$~K. This cooling
limit is associated with the magnetic entropy freezing at the
transition into an ordered state. When the field is further
decreased, a weak temperature increase is observed. Two
temperature plateaus (shown by the arrows in Fig.~\ref{fig3}) are
clearly seen in this part of the curve corresponding to the phase
transitions at $T_{N1}\simeq 1$ K and $T_{N2}\simeq 0.75$ K
\cite{ramirez02,bonville,oleg03}.

It is interesting to compare the above measurements to the
magnetocaloric effect in another frustrated spin system \GGG. The
data on \GGG\ have been obtained on a sintered powder sample (mass
4.05 g) using an experimental technique similar to that described
above. Due to the large heat capacity of the sample, no correction
for parasitic heat leaks is required in this case. The measured
adiabatic demagnetization curves for \GGG\ are very similar to the
previously published results for oriented single
crystals\cite{fisher} in the temperature range of overlap. Our
results are compared to \GTO\ data in Fig.~\ref{fig3} on a field
scale normalized to the corresponding values of $H_{\rm sat}$:
15~kOe and 70 kOe, respectively. The magnetocaloric effect in the
\GGG\ sample is qualitatively similar to what is observed in \GTO,
but occurs in a different temperature range due to a smaller
exchange constant. In a demagnetization process of \GGG\ starting
from $H\simeq 2H_{\rm sat}$ (28 kOe) at $T=0.5$~K the temperature
decreases by a factor of 5 reaching its minimum value of 0.1~K in
the vicinity of the saturation field after which it remains
practically constant.

\section{Discussion}

Our experimental results demonstrate the decisive role of the
frustration in producing large adiabatic temperature changes in
the two materials. In typical nonfrustrated antiferromagnets,
magnetocaloric effect in the vicinity of field induced transitions
is very small. For example, adiabatic $\Delta T_S$ does not exceed
0.05~K in CsNi$_{0.9}$Fe$_{0.1}$Cl$_3$i for a similar range of
fields and temperatures.\cite{enderle} For a quantitative
description of the above data we have performed a series of Monte
Carlo (MC) simulations on classical antiferromagnetic models with
only nearest-neighbor exchange. A detailed account of the
application of the MC technique to model adiabatic processes is
given in Ref.~\onlinecite{mzh03}. Quantum spins $S=7/2$ are
replaced by classical $n$-vectors ($|{\bf n}|=1$), an
approximation which is valid for high enough temperatures $T\agt
JS$. The only parameter required to compare the MC simulations to
the experimental data is the exchange constant $J$, which is
estimated from values of the saturation field or the Curie-Weiss
temperature. Expressions for these quantities for nearest-neighbor
antiferromagnets on pyrochlore and garnet lattices are
\begin{eqnarray}
g\mu_B H_{\rm sat} & = & 8JS\ ,\ \ \ k_B\theta_{CW} = 2JS(S+1)\ ,
\nonumber \\ g\mu_B H_{\rm sat} & = & 6JS\ ,\ \ \ k_B\theta_{CW} =
{\textstyle \frac{4}{3}}JS(S+1)\ ,
\end{eqnarray}
respectively (see, {\it e.g.}, Ref.~\onlinecite{mzh03}).
Corresponding values for both compounds have been taken
from previous works \cite{raju,ramirez02,bonville,ggg} and are
summarized in Table \ref{tab1}. For both materials, the exchange
constants derived from $H_{\rm sat}$ and $\theta_{CW}$ agree
to within an accuracy of $\sim 10$~\%.
The remaining difference may be attributed to weaker dipolar and other
interactions, which are responsible for the eventual ordering
in \GTO\ below $T_{N1}$ and above $H=5$~kOe in \GGG.

\begin{table}[h]
\caption{Experimental values $H_{\rm sat}$ and $\theta_{CW}$ and
the estimations of the exchange constants for \GTO\ and \GGG.}
\begin{tabular}{|c|c|c|c|c|}  \hline
&~$H_{\rm sat}$~ & ~$\theta_{CW}~$ & $~J(H_{\rm sat})~$ &
~$J(\theta_{CW})~$ \\ \hline \GTO\ & ~70 kOe~ & ~9.5 K~ & ~0.33
K~& ~0.30 K~ \\ \GGG\ & 15 kOe & 2.3 K & 0.1 K & 0.11 K \\ \hline
\end{tabular}
\label{tab1}
\end{table}

The results of the MC simulations for \GTO\ with $J=0.3$~K are
indicated by the solid lines in Figs.~\ref{fig2} and \ref{fig3}.
For the scans starting at temperatures above 10~K one can not
neglect the phonon contribution to the total heat capacity of the
sample. The influence of lattice heat on the demagnetization
process can be taken into account using Eq.~(\ref{balance}), which
yields the following relationship between the ideal temperature
variation $\Delta T_S$ and its observed value $\Delta T_{\rm
real}$:

\begin{equation}
\Delta T_S \simeq \Delta T_{\rm real} (1+C_{\rm ph}/C),
\end{equation}

\noindent where $C_{\rm ph}$ is the phonon part of the specific
heat, $C$ is its magnetic part. Using the results shown in Fig.
\ref{fig1}, one can approximately correct the $T_S(H)$ curves for
the temperature dependent phonon contribution (see dotted lines on
Fig. \ref{fig2}). With this correction, the MC calculations
demonstrate very good agreement with the experiment in the whole
temperature range at fields above $H_{\rm sat}$. Since a
nearest-neighbor Heisenberg pyrochlore antiferromagnet does not
order, the MC simulations cannot describe satisfactorily the observed
behavior below $H_{\rm sat}$ at $T<1$~K. Besides, this is also the
limit of the validity of the classical approximation: $JS\simeq
1$~K. The temperature increase at $H<H_{\rm sat}$ observed in
\GTO\ (Fig.~\ref{fig3}) may be attributed to the reopening of a
gap in the excitation spectrum below $H_{\rm sat}$ caused by
anisotropic interactions. The theoretical dependence obtained for
\GGG\ (lower solid line in Fig.~\ref{fig3}) with the exchange
constant $J=0.11$~K also fits
 the high field part of the experimental adiabatic curve very well.

\begin{figure}[t]
\centerline{\includegraphics[width=\columnwidth]{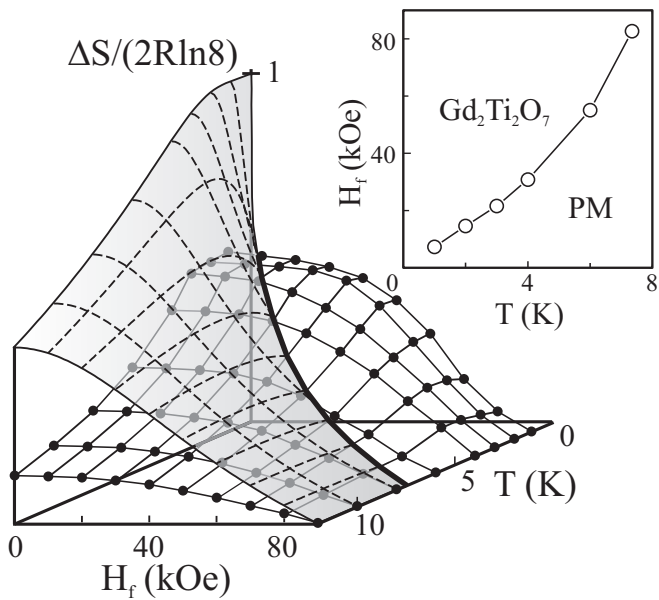}}
\caption{The molar entropy change of \GTO\ under isothermal
demagnetization from $H_i=90$~kOe to $H_f$ at different
temperatures. Entropy variations of an ideal $S=7/2$ paramagnet
under the same conditions are shown by the dashed lines. The inset
represents the critical intersection line of the two surfaces.}
\label{cool}
\end{figure}

Finally, we estimate the cooling power of the demagnetization
process using the measured adiabatic $T_S(H)$ curves along with
the specific heat data $C(T)$ obtained at high fields. The amount
of heat absorbed by a magnetic material during isothermal
demagnetization is related to the entropy change $\Delta Q = T
\Delta S|_{H_i}^{H_f}$. Consider an adiabatic demagnetization
curve, which starts at $(H_i,T_i)$ and ends at $(H_f,T_f)$. The
entropy along the curve $T_S(H)$ remains constant, so we can
relate the entropy variations at constant temperature and in
constant field:

\begin{equation}
\Delta S(T_f)\left|_{H_i}^{H_f}\right. = \Delta
S(H_i)\left|_{T_f}^{T_i}\right. = \int_{T_f}^{T_i}C(T)/T\: dT \ .
\label{cooling}
\end{equation}

The entropy changes $\Delta S$ of \GTO\ under the isothermal
demagnetization at various temperatures from the initial field
$H_i=90$ kOe as a function of the final field $H_f$ are presented
by closed circles in Fig.~\ref{cool}. The values of $\Delta S$ are
calculated from our experimental data using the above equation,
while the real entropy of the system remains undefined. Two
features should be underlined: (i) about one half of the total
magnetic entropy $2R\ln 8$ remains in the system even at
temperatures very close to the transition into an ordered state at
$T_{N1}=1$ K; (ii) the largest entropy change $\Delta S$ and
consequently, a heat absorption $\Delta Q$ occurs in a high field
region above $H_{\rm sat}$. This differs significantly from the
behavior of an ideal paramagnet at low temperatures (shown by grey
surface on Fig.~\ref{cool}), which releases a significant part of
its entropy only if demagnetized to $H_f\ll H_i$. This figure
demonstrates that \GTO\ has a considerable advantage below 5~K in
the {\it high field} cooling power over conventional paramagnets.
The boundary of the field-temperature area where the pyrochlore is
advantageous over an ideal $S=7/2$ paramagnet obtained as an
intersection of the two sheets (the bold line) is shown in the
inset to Fig. \ref{cool}. For comparison, we performed similar MC
simulations of a nearest-neighbor antiferromagnet (not shown in
the graph). Classical models cannot give correct values for the
total entropy of a quantum spin system. Nevertheless, the
variation of entropy is reproduced with a remarkable accuracy. For
example, the calculated values of $\Delta S$ at $T=4.2$ K for
demagnetization from $H_i=90$ kOe to zero field and to $H_f=40$
kOe are 0.52 and 0.40 of the total magnetic entropy respectively.
The difference from the experiment does not exceed 20\%, which is
due to the large $S=7/2$ spin of the Gd$^{3+}$ ions.

The cooling power $\Delta Q$ has a maximum around 4~K reaching
30~J/mole Gd. Such an amount of heat corresponds, for comparison,
to the evaporation heat of approximately one mole of liquid $^3$He
at $T=3$~K. The garnet compound \GGG\ is also advantageous as a
refrigerant material over paramagnetic salts, but at lower fields
and temperatures. We suggest that a combination of the two
compounds might become the basis for a inexpensive two-stage
adiabatic demagnetization refrigerator suitable for effective
cooling from $T\sim 10$~K down to 100 mK range in a single field
sweep.

\section{Conclusions}

In conclusion, a large magnetocaloric effect is observed in the
frustrated pyrochlore antiferromagnet \GTO\ in agreement with
recent theoretical predictions.\cite{mzh03,mzh04} This observation
points at the presence of a macroscopic number of local low-energy
excitations in the vicinity of $H_{\rm sat}$. Such modes can be
directly probed in quasielastic neutron scattering measurements
analogously to the experiment on ZnCr$_2$O$_4$.\cite{broholm}
These excitations may also be responsible for the nonfrozen spin
dynamics below $T_{N1}$ observed in \GTO\ by muon spin relaxation
measurements. \cite{yaouanc} A comparison between our experimental
data and classical MC simulations shows that this numerical
technique can semiquantitatively predict the magnetocaloric
properties of real rare-earth materials with large (semiclassical)
magnetic moments, which have been described so far only in the
molecular field approximation.

\acknowledgments

The authors are grateful to V. I. Marchenko and M.~R.~Lees for
valuable discussions. This work is supported by RFBR grant
04-02-17294 and by the RF President Program. S.S.S. is also
grateful to the National Science Support Foundation for the
financial help.

\end{document}